\documentclass[aps,prl,floatfix,superscriptaddress,twocolumn,footinbib]{revtex4-2}

\usepackage[T1]{fontenc}

\usepackage[utf8]{inputenc}
\usepackage{xcolor}
\usepackage{mathrsfs}
\usepackage{amsmath}
\usepackage{amssymb}
\usepackage{makeidx}
\usepackage{amsfonts,amssymb}
\makeindex
\usepackage{graphicx}

\makeatletter

\usepackage{hyperref}
\hypersetup{colorlinks = true,allcolors = blue}
\usepackage{times}

\makeatother

\usepackage{babel}
\begin{document}

\title{Intrinsic Quantum Valley Hall Effect in Monolayer System}

\author{Yu-Hao Wan}
\affiliation{International Center for Quantum Materials and School of Physics,
Peking University, Beijing 100871, China}
\author{Cheng-Ming Miao}
\affiliation{International Center for Quantum Materials and School of Physics,
Peking University, Beijing 100871, China}
\author{Yu-Cheng Zhuang}
\affiliation{International Center for Quantum Materials and School of Physics,
Peking University, Beijing 100871, China}
\author{Qing-Feng Sun}
\email[Corresponding author: ]{sunqf@pku.edu.cn.}
\affiliation{International Center for Quantum Materials and School of Physics, Peking University, Beijing 100871, China}
\affiliation{Hefei National Laboratory, Hefei 230088, China}

\begin{abstract}
We report an intrinsic quantum valley Hall (QVH) phase in monolayer honeycomb lattices that does not rely on real-space domain walls. A sublattice-dependent next-nearest-neighbor hopping generates a momentum-dependent mass that stabilizes valley-polarized edge states within a uniform bulk. We show that this phase admits a simple global geometric characterization: the Hamiltonian vector field $\mathbf{d}(\mathbf{k})$ forms a bowl-shaped surface that encloses the origin in parameter space. In this geometry, band-inversion surfaces (BIS) appear as closed loops that wind around the valley charges at $K$ and $K'$, providing a compact topological criterion for the emergence of momentum-separated valley edge modes. Transport simulations confirm a quantized, disorder-robust valley-polarized edge transport. Because the mechanism is formulated purely in terms of the $\mathbf{d}(\mathbf{k})$ texture, it offers a lattice-agnostic design principle for intrinsic valleytronics in electronic, photonic and acoustic metamaterials.
\end{abstract}

\maketitle
\textit{Introduction}-
Valley is an internal degree of freedom of electrons in monolayer honeycomb lattices such as graphene, arising from the inequivalent $K$ and $K^{\prime}$ points in the Brillouin zone and playing a role analogous to spin. This concept has laid the foundation for valleytronics \cite{schaibley_valleytronics_2016}, a field dedicated to utilizing the valley index for information encoding and processing. Central to this field is the quantum valley Hall (QVH) effect \cite{PhysRevLett.102.096801,PhysRevLett.133.196603}, which hosts valley-polarized edge states that are robust against scattering from smooth potentials, and has significant application prospects in valleytronics \cite{schaibley_valleytronics_2016,PhysRevLett.132.166601}, quantum computing \cite{Funayama-AS2025}, and topological photonics \cite{lu_observation_2017, dong_valley_2017, PhysRevLett.120.116802, barczyk_observation_2024}.
To realize the QVH effect, strategies have predominantly relied on extrinsic engineering. One common approach in monolayer honeycomb lattices is to engineer domain walls that confine valley modes at their interfaces, by creating adjacent regions with opposite signs of an A/B-staggered sublattice potential so that the interface acts as a topological domain wall \cite{PhysRevLett.127.116402,PhysRevLett.101.087204,PhysRevLett.121.156801}. Another utilizes strain engineering to break the lattice periodicity \cite{PhysRevLett.129.076802}, generating pseudo-magnetic fields that drive valley transport \cite{guinea_energy_2010,low_strain-induced_2010,ZYCPRB2022,addchang}. Beyond monolayers, gated bilayer and few-layer graphene provide a solid-state platform where the interplay of perpendicular electric fields and stacking textures generates QVH phases with tunable valley kink and edge channels \cite{PNAS, PhysRevB.84.075418,ju_topological_2015, li_gate-controlled_2016, yin_direct_2016, doi:10.1126/science.adj3742}.

Here lies a fundamental irony: while the valley concept itself originated in monolayer honeycomb lattices, these systems surprisingly lack an intrinsic QVH phase. A truly intrinsic realization—one arising naturally in the bulk without external domain boundaries, strain gradients, or multi-layer stacking—remains elusive in the fundamental monolayer system. Furthermore, the QVH effect currently lacks a unified global bulk-boundary correspondence \cite{PhysRevB.82.245404}. The prevailing description relies on the valley Chern number, a quantity that is somewhat artificially defined by restricting the integration to a local region near the Dirac points based on low-energy effective models \cite{XiaoPRL2007,PhysRevB.97.174301,PhysRevLett.133.196603}. This stands in sharp contrast to Chern insulators, where the topological invariant arises from a global integration over the full Brillouin zone \cite{asboth_short_2016,wan2025classification}, leaving the establishment of a global topological framework for QVH an open challenge.

In this Letter, we bridge this gap by introducing a mechanism driven solely by sublattice-dependent next-nearest-neighbor hopping amplitudes. This intrinsic modification induces a momentum-dependent mass term that stabilizes a QVH phase and naturally generates valley-polarized edge states within the monolayer system, effectively shifting the paradigm from extrinsic engineering to intrinsic band-geometry control. Crucially, we provide a global geometric interpretation: the Hamiltonian vector field $\mathbf{d}(\mathbf{k})$ forms a "bowl-shaped" surface enclosing the origin, offering a new and rigorous topological classification that parallels Chern insulators but is tailored to time-reversal-symmetric systems. This topology could also extend beyond electronics, offering a universal framework for controlling valley transport in photonic, phononic, and mechanical metamaterials.

\begin{figure}[b]
\raggedright  %
\includegraphics[width=1\linewidth]{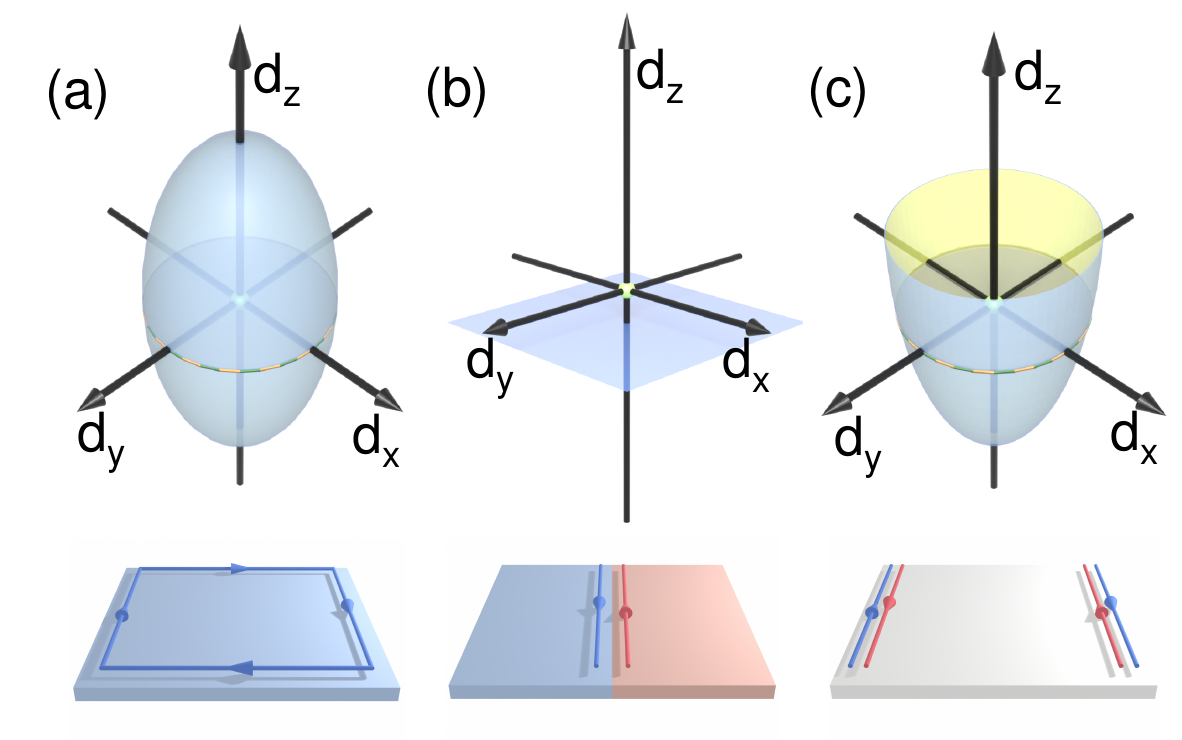}
\caption{\label{fig:1}
(a) A typical Chern-insulator phase:
Top: $\mathbf{d}_\mathbf{k}$ surface fully encloses the origin ($C=1$).
Bottom: Corresponding chiral edge modes appear at the boundary.
(b) Conventional valley Hall phase in staggered graphene:
Top: Flat $\mathbf{d}_\mathbf{k}$ plane that does not enclose the origin.
Bottom: Valley edge modes arise only at domain walls.
(c) Curvature-driven bowl-shaped QVH phase which is raised in this word:
Top: $\mathbf{d}_\mathbf{k}$ surface bends downward and encloses the origin.
Bottom: Momentum-separated edge states emerge at the boundary.
}
\end{figure}

\textit{Model and Geometric Mechanism}. —
We first introduce the general framework of a two-band Hamiltonian
$H(\mathbf{k})=\mathbf{d}(\mathbf{k})\cdot$$\boldsymbol{\sigma}$,
where $\mathbf{d}(\mathbf{k})=\left(d_{x}(\mathbf{k}),d_{y}(\mathbf{k}),d_{z}(\mathbf{k})\right)$ defines
a continuous mapping from the Brillouin zone $T^{2}$ to $\mathbb{R}^3$ \cite{aidelsburger_artificial_2018}.
The topology of the insulating phase is encoded in the geometry of
the $\mathbf{d}_\mathbf{k}$ surface traced by $\mathbf{d}(\mathbf{k})$ in this three-dimensional
parameter space. A paradigmatic example is the Chern insulator: when the $\mathbf{d}_\mathbf{k}$ surface forms a closed sphere enclosing the origin, the normalized vector field $\hat{\mathbf{d}}(\mathbf{k})$ realizes a mapping $T^2 \to S^2$ with a nonzero Chern number [See Fig. \ref{fig:1}(a)]. In this geometric picture,
the Chern number counts how many times the closed surface $\mathbf{d}(\mathbf{k})$
wraps around the origin, with the wrapping orientation determining
the sign of $C$ \cite{wan2025classification, bansil_colloquium_2016}. A nonzero $C$ implies that the bulk bands carry a nontrivial topology,
manifested by $|C|$ chiral edge modes at the boundary.%

By contrast, Fig. \ref{fig:1}(b) shows the well-known QVH effect in gapped graphene with a staggered sublattice potential. Here, the sublattice asymmetry creates a constant mass term ($d_z = \Delta_{AB}$) that is independent of momentum. Consequently, the $\mathbf{d}_\mathbf{k}$ surface collapses into a flat plane. Since this plane does not enclose the origin, the Chern number vanishes. As a result, a uniform sample is topologically trivial and hosts no edge modes \cite{gao_topologically_2018}. Valley Hall edge modes emerge only at domain walls where the sublattice potential reverses sign—that is, at interfaces between regions of opposite staggering.

Finally, Fig. \ref{fig:1}(c) illustrates the new type of topology
proposed in this work, where the $\mathbf{d}_\mathbf{k}$ surface takes a bowl-like
shape that encloses the origin. Systems with such geometry naturally host valley-polarized edge states, whose number corresponds to the folding degree of the $\mathbf{d}_\mathbf{k}$ surface. Unlike conventional QVH realizations that rely on domain-wall engineering, this topology arises intrinsically from the global geometry of the $\mathbf{d}_\mathbf{k}$ mapping-its bowl-shaped wrapping around the origin directly signals a nontrivial QVH configuration.

\begin{figure}
\begin{centering}
\includegraphics[scale=0.12]{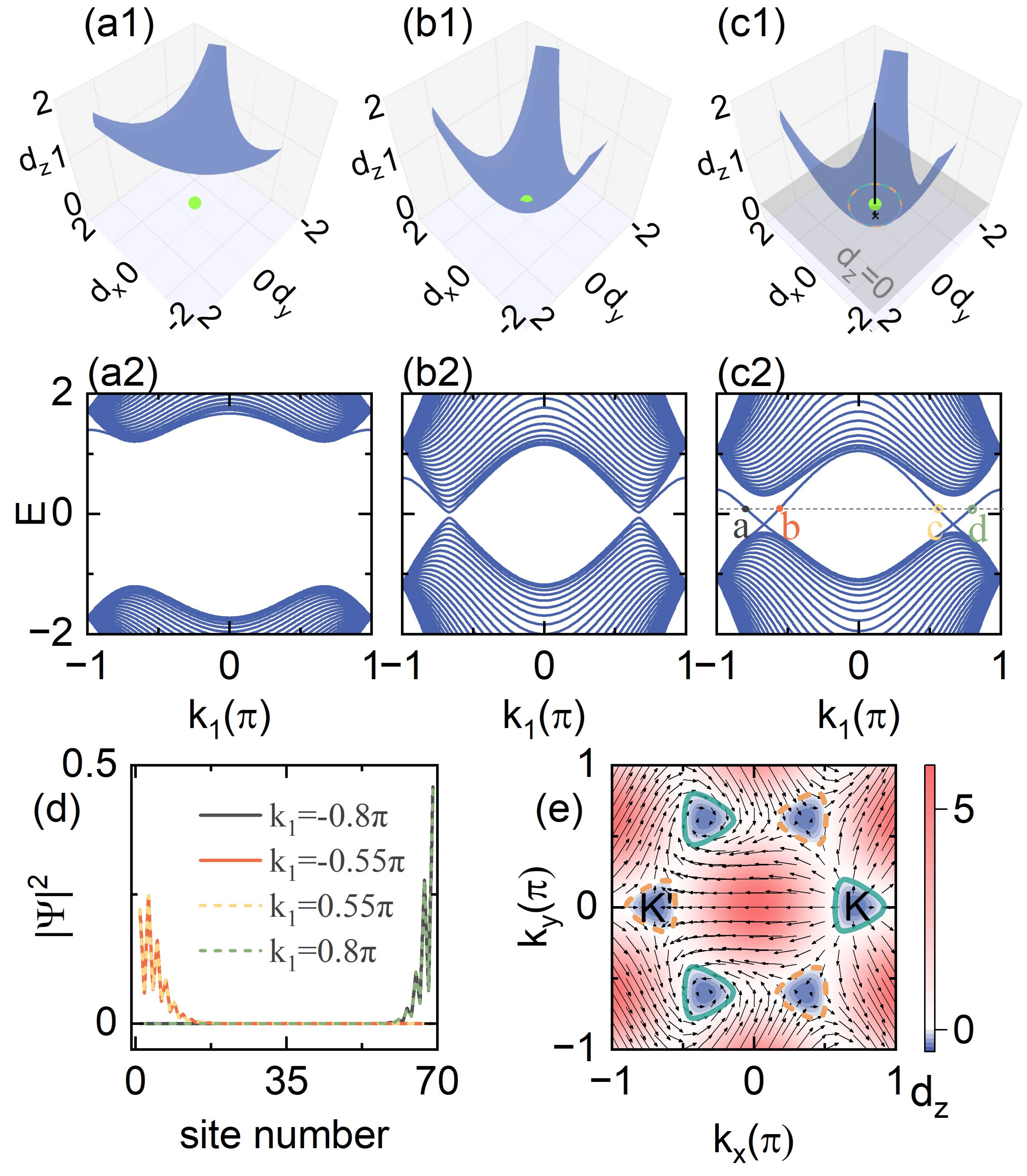}
\par\end{centering}
\caption{\label{fig:2} (a1-c1) Evolution of the $\mathbf{d}_\mathbf{k}$ surface with
increasing next-nearest-neighbor hopping $t_{2}=0.2,0.6,0.7$.
The orange and green loops in (c1) denote the BIS, the green sphere in (a1-c1) marks the origin, and the crosses ($\times$) in (c1) indicate the topological charges at valleys $K$ and $K'$.
(a2-c2) Corresponding nanoribbon spectra for a ribbon periodic along the zigzag
direction and finite across the transverse width ($W=30a$).
Both zigzag and bearded edge terminations are employed in the calculation.
(d) Real-space probability distributions
of the eigenstates marked by colored dots in (c2).
(e) Distribution of the $\mathbf{d}(\mathbf{k})$ vector over $\left(k_{x},k_{y}\right)\in[-\pi,\pi]$.
The BIS, shown as orange and green loops, form closed contours around
the inequivalent valleys $K$ and $K^{\prime}$.
Although the two BIS appear overlapped in the $\mathbf{d}_\mathbf{k}$ surface of
(c1), they are well separated in momentum space and individually encircle
the valley charges at $K$ and $K^{\prime}$. %
All calculations use $t=1$ and $\Delta_{AB}=1.8$.}
\end{figure}

To illustrate this geometric mechanism, we consider a honeycomb lattice described by
\begin{equation}
H=-t \sum_{\langle i, j\rangle} c_i^{\dagger} c_j+\sum_{\langle\langle i, j\rangle\rangle} t_2^{s_i} c_i^{\dagger} c_j+\sum_i s_i {\Delta_{A B}} c_i^{\dagger} c_i, \label{Eq1}
\end{equation}
where $c_i^{\dagger}\left(c_i\right)$ creates (annihilates) an electron on site $i$, $t$ is the nearest-neighbor hopping between A and B sublattices, and $t_2^{s_i}=s_i t_2\left(s_i=+1\right.$ for $\mathrm{A}$ and $-1$ for B$)$ denotes the sublattice-dependent next-nearest-neighbor hopping.
The last term introduces a staggered sublattice potential $\pm \Delta_{A B}$. This minimal model contains all ingredients necessary for realizing the intrinsic QVH phase.
Upon Fourier transforming to momentum space, the Hamiltonian acquires the form $H(\mathbf{k})=\mathbf{d}(\mathbf{k}) \cdot \boldsymbol{\sigma}$, %
where the $\mathbf{d}_\mathbf{k}$ surface naturally visualizes the band geometry and topology (See Sec. S1 of the Supplemental Material \cite{seesupplemental}).

Figure \ref{fig:2} summarizes how the system evolves with increasing next-nearest-neighbor
hopping $t_2$. Panels (a1-c1) show the deformation of the $\mathbf{d}_\mathbf{k}$ surface, while panels (a2-c2) present the corresponding ribbon spectra for a geometry that is periodic along the zigzag direction and finite in the transverse direction. %
For small $t_{2}$, the normal component $d_{z}$ is nearly momentum
independent, so the $\mathbf{d}_\mathbf{k}$ surface appears as a slightly tilted
plane. As shown in Fig. \ref{fig:2}(a1), the origin is not enclosed and the system remains
topologically trivial: a uniform sample carries no edge states. The
corresponding ribbon spectrum in Fig. \ref{fig:2}(a2) shows a full bulk gap without
edge modes.

Increasing $t_{2}$ continuously bends the $\mathbf{d}_\mathbf{k}$ surface into a
bowl. At a critical $t_{2}=\Delta_{AB}/3$ \cite{seesupplemental}, the bottom
of the bowl touches the origin, as illustrated in Fig. \ref{fig:2}(b1).
In this case, the bulk gap closes [See Fig. \ref{fig:2}(b2)]. 
Further increasing $t_{2}$, the bottom of the $\mathbf{d}_\mathbf{k}$ surface
is lower than the origin, leading to the origin enclosed by the $\mathbf{d}_\mathbf{k}$
surface [see Fig. \ref{fig:2}(c1)]. As a result, the bulk gap reopens and valley edge states
emerge inside the bulk gap. 
The corresponding ribbon spectrum in Fig. \ref{fig:2}(c2) clearly 
displays edge branches near the $K$ and $K^{\prime}$ valleys. 
From these in-gap states, we select two pairs of modes with opposite 
Fermi velocities at $K$ and $K^{\prime}$; their real-space wavefunction distributions
are shown by corresponding colored lines in Fig. \ref{fig:2}(d). 
One sees that the two valleys host counterpropagating edge channels
localized on opposite sides of the ribbon, revealing their opposite
chiralities.

Moreover, these valley Hall edge states are insensitive to the specific boundary termination:
as shown in the Supplemental Material \cite{seesupplemental}, they persist for both zigzag and bearded edges, underscoring their bulk topological origin. 
These momentum-separated, valley-polarized edge states constitute the characteristic transport signature of the QVH phase.
Importantly, this geometric mechanism is not limited to the honeycomb lattice. 
A square-lattice analogue with a similar bowl-shaped $\mathbf{d}_{\mathbf{k}}$ surface 
is shown to host intrinsic valley Hall edge states, as further demonstrated in the End Matter, underscoring the universality of the geometric mechanism.

\textit{Topological Origin and BIS.} —
To elucidate the microscopic origin of the valley topology, we examine
the full $\mathbf{d}(\mathbf{k})$ field in the Brillouin zone. 
Figure \ref{fig:2}(e), corresponding to the parameters of Fig.
\ref{fig:2}(c), displays the distribution
of $\mathbf{d}(\mathbf{k})$ over $\left(k_{x},k_{y}\right)\in[-\pi,\pi]$,
providing a direct visualization of the underlying topological structure.%
Two topological features are immediately apparent in Fig. \ref{fig:2}(e).
First, topological charges exist at isolated points
where $d_{x}(\mathbf{k})=d_{y}(\mathbf{k})=0$;
in the present model, these coincide with the two inequivalent valleys
$K$ and $K^{\prime}$. Each charge acts as a vortex center of the
in-plane vector $\mathbf{d}_{\perp}=\left(d_{x},d_{y}\right)$, and
the two vortices carry opposite chiralities.
Second, one can identify band-inversion surfaces (BIS),
defined by $d_{z}(\mathbf{k})=0$ \cite{yu2021quantum,song2019observation,wan2025classification,wan2025interplay,yi2019observing,hu2020topological}. As shown by the orange and green
loops in Fig. \ref{fig:2}(e), the BIS loops encircle the two valley charges
separately-each BIS winds around one vortex of $\mathbf{d}_{\perp}$.

This BIS-charge configuration directly corresponds to the bowl-shaped 
$\mathbf{d}_\mathbf{k}$ geometry enclosing the origin, as shown in Fig. \ref{fig:2}(c1).
In this picture, the topological charges are located at the intersection 
points between the $\mathbf{d}_\mathbf{k}$ surface and the $d_z$ axis 
at the bottom of the bowl [marked by $\times$ in Fig. \ref{fig:2}(c1)], 
while the BIS loops arise precisely where the concave $\mathbf{d}_k$ surface 
crosses the plane $d_z=0$ [gray plane in Fig. \ref{fig:2}(c1)]. 
Although the two BIS loops appear to overlap on the three-dimensional
$\mathbf{d}_\mathbf{k}$ surface, they are well separated in momentum space 
[see Fig. \ref{fig:2}(e)], each enclosing a distinct valley charge. 
The bowl enclosing the origin thus guarantees that both BIS loops wrap 
nontrivially around their respective charges. 
Such BIS-charge windings constitute the essential
topological structure of the QVH phase. 
Previous studies have suggested a correlation between the BIS-charge windings 
and the Chern number \cite{Zhang2018ask,hu2020topological,PhysRevResearch.2.023043,wan2025classification}. 
Here, this BIS-charge correspondence provides a direct visualization
of the valley-resolved topology: the local winding of $\mathbf{d}_{\perp}$
along each BIS defines an integer valley Chern number $C_K=+1$ and $C_{K^{\prime}}=-1$, 
with the opposite signs at $K$ and $K^{\prime}$ guaranteeing counter-propagating valley edge states.

\begin{figure}
\begin{centering}
\includegraphics[scale=0.35]{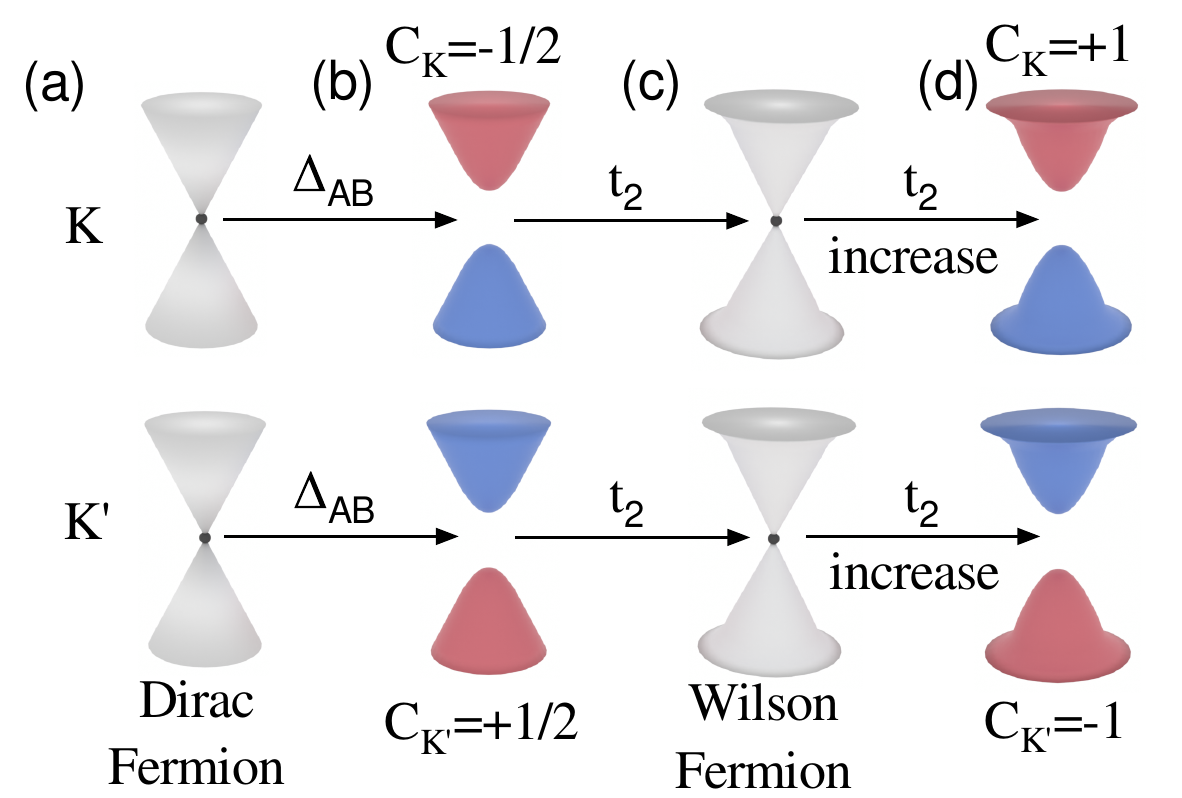}
\par\end{centering}
\caption{\label{fig:3} (a-d) Schematic evolution from Dirac (a) to Wilson (b) fermions 
at the two valleys.
A sublattice potential $\Delta_{A B}$ opens opposite gaps at $K$ and $K^{\prime}$, 
yielding half-quantized valley Chern numbers (b). 
Increasing the next-nearest-neighbor hopping $t_2$ introduces a Wilson mass, which
leads to the closing and reopening of the bulk gap (b-d), thereby transforming each Dirac cone into a massive Wilson fermion with integer valley Chern numbers $C_K=+1$ and $C_{K^{\prime}}=-1$.}
\end{figure}

Next, let us understand how valley topology emerges from a low-energy perspective.
Figure \ref{fig:3}(a) shows the gapless Dirac cones
at $K$ and $K^{\prime}$ in pristine graphene-massless Dirac fermions.
Introducing a staggered sublattice potential [Fig. \ref{fig:3}(b)] just opens
a mass gap at each valley and endows every Dirac fermion with a constant
Dirac mass $M_{K/K^{\prime}}=\pm\Delta_{AB}$. %
Each valley thus hosts a massive
Dirac cone carrying a half-quantized valley Chern number $\nu_{K/K^{\prime}}=\mp\frac{1}{2}$ \cite{XiaoPRL2007}.
Valley edge states must require a sign reversal of the mass at a domain
wall, as in the Jackiw-Rebbi mechanism\cite{jackiw1976solitonsa},
necessitating a heterostructure interface. 
Adding the sublattice-dependent next-nearest-neighbor hopping $t_2$
bends the $\mathbf{d}_\mathbf{k}$ surface into a bowl. 
From a low-energy perspective, this corresponds to introducing a momentum-dependent
Wilson mass, and simultaneously renormalizes the Dirac mass 
(see Sec. S3 of the Supplemental Material \cite{seesupplemental}).
As the system undergoes a gap closing and reopening with the 
increase of $t_2$ [Fig. \ref{fig:3}(c)], the interplay between 
the Dirac and Wilson masses drives each valley from half-integer 
Chern number to an integer valley Chern number [Fig. \ref{fig:3}(d)],
further giving rise to intrinsic, domain-wall-free valley Hall edge states.
Such a novel configuration has not been noticed in previous studies 
and we also demonstrate that it is a direct manifestation of the bulk topology.

\begin{figure}
\begin{centering}
\includegraphics[scale=0.35]{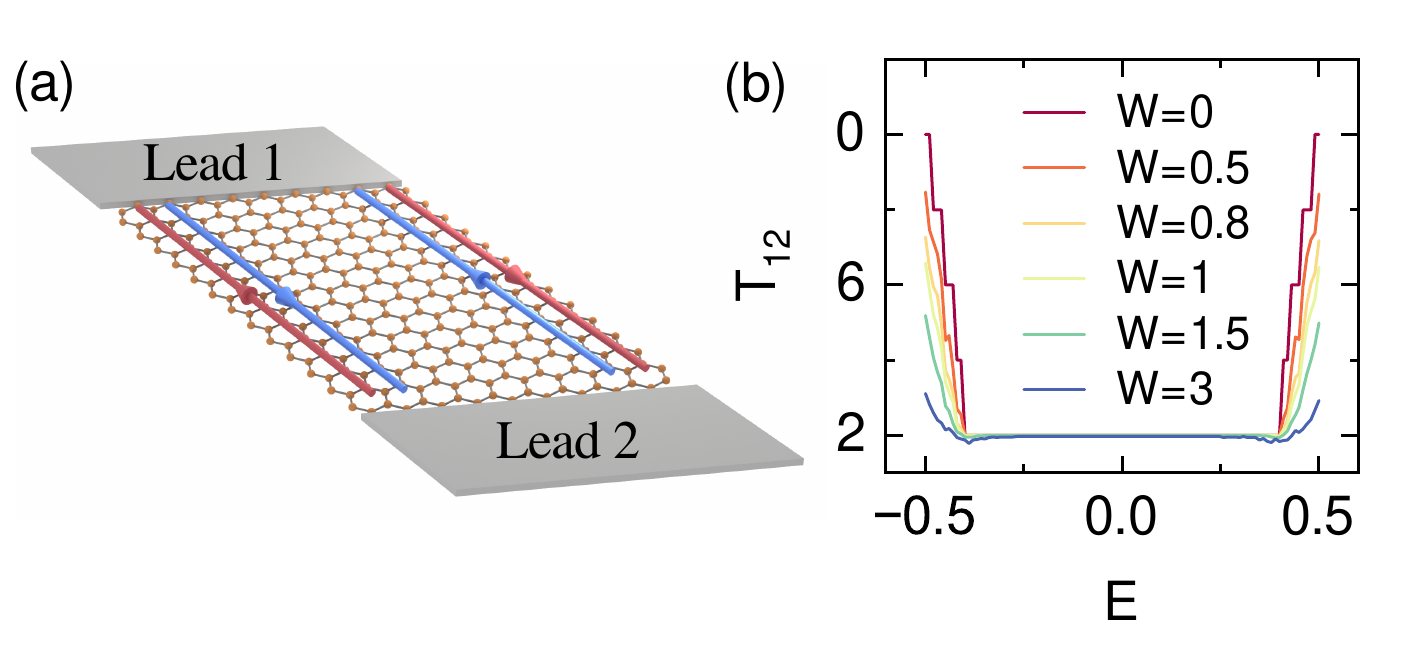}
\par\end{centering}
\caption{\label{fig:4} (a) Schematic of the two-terminal transport setup with
zigzag boundaries. Red and blue arrows indicate the counter-propagating
valley-polarized edge channels associated with $K$ and $K'$ valleys. (b) Energy
dependence of the transmission probability between terminals in the
presence of disorder with varying strengths. }
\end{figure}

\textit{ Robust Valley Transport}.\textemdash
Having established the bowl-BIS bulk bands-QVH edge modes correspondence, we now turn to their transport manifestations.
We consider a two-terminal zigzag ribbon as sketched in Fig. \ref{fig:4}(a).
The valley-polarized channels from $K$ and $K^{\prime}$ counter-propagate along
opposite edges (red and blue arrows), forming momentum-separated valley currents. To probe robustness, we add long-range scalar disorder in the
scattering region as an onsite term $\epsilon_{i}\sigma_{0}$ and
compute the energy-resolved transmission $T_{12}(E)$ %
between the
two terminals, which quantifies the transmission probability from
lead 1 to lead 2. The disorder field is generated by convolving independent
random onsite potentials variables $\tilde{\epsilon}_{j}\in[-W/2,W/2]$
with a normalized Gaussian kernel of width $\eta$ \cite{wan2025interplay,wan2025classification},
\[
\epsilon_{i}=\frac{\sum_{j}\tilde{\epsilon}_{j}e^{-|\mathbf{r}_{i}-\mathbf{r}_{j}|^{2}/(2\eta^{2})}}{\sqrt{\sum_{j}e^{-|\mathbf{r}_{i}-\mathbf{r}_{j}|^{2}/(2\eta^{2})}}},
\]
truncated at $3\eta$ in numerics. We take $\eta=1.5a$ (lattice constant
$a=1$) and keep the leads clean. The transmission is obtained from $T_{12}(E)=\mathrm{Tr}\left[\Gamma_{2}G^{R}\Gamma_{1}G^{A}\right]$ \cite{zhou_transport_2022,wan_altermagnetism-induced_2025,wan_quarter-quantized_2024}, %
and averaged over $N=100$ disorder realizations for each W (details in Sec. S4 \cite{seesupplemental}).

Figure \ref{fig:4}(b) shows transmission $T_{12}$ as a function of energy $E$ 
for different disorder strength $W$. 
In the absence of disorder ($W=0$), $T_{12}$ manifests quantized plateau 
with a value of $2$ within the bulk gap and high values outside.
This QVH plateau remains quantized even under strong disorder ($W=3$), 
whereas the high values of $T_{12}$ from bulk contribution rapidly
decreases as the disorder strength increases. 
This robustness reflects the BIS-protected, momentum-separated nature of
the $K/K'$ channels together with the fact that long-range disorder
inefficiently induces intervalley scattering.
Together, these results demonstrate domain-wall-free quantized valley transport that is inherently
robust against long-range disorder, directly confirming the bulk-topological
origin of the QVH edge modes.

$\textit{Experiment Realization}$.\textemdash
Let's discuss the experimental feasibility.
In our model in Eq. (\ref{Eq1}), the bowl-shaped $\mathbf{d}_\mathbf{k}$
surface and the resulting QVH topology originate from a sublattice-dependent
opposite next-nearest-neighbor hopping
$t_{2}^{AA}=+t_{2}$ and $t_{2}^{BB}=-t_{2}$,
which introduces a momentum-dependent mass term 
and bends the $\mathbf{d}_\mathbf{k}$ surface into a bowl.
More generally, any finite asymmetry
$t_{2}^{AA}\neq t_{2}^{BB}$ can be decomposed as $t_{2}^{0}=\frac{1}{2}\left(t_{2}^{AA}+t_{2}^{BB}\right),\quad t_{2}^{z}=\frac{1}{2}\left(t_{2}^{AA}-t_{2}^{BB}\right)$,
where $t_{2}^{0}$ multiplies $\sigma_{0}$ (topologically inert)
while $t_{2}^{z}$ multiplies $\sigma_{z}$ and directly controls
the bowl geometry. %
We verified numerically (see Sec. S5 \cite{seesupplemental}) that finite $t_{2}^{z}$ reproduces
the same valley-polarized edge states as the idealized choice $t_{2}^{AA}=-t_{2}^{BB}$,
confirming that the topological diagnosis by the valley Chern number
$C_{K/K^{\prime}}$ and the corresponding QVH edge modes remain
unchanged.

From a materials perspective, such a sublattice-dependent next-nearest-neighbor hopping
arises naturally when the two sublattices host inequivalent orbitals
or local bonding environments, leading to unequal overlap integrals.
This condition can be engineered in various hexagonal systems where
$\mathrm{A}/\mathrm{B}$ asymmetry is intrinsic or tunable-such as
van der Waals heterostructures and monolayers with valley physics
at $K/K^{\prime}$ (e.g., hBN- or TMD-based platforms), through substrate \cite{PhysRevB.88.085433,PhysRevB.92.205108}.
These routes provide realistic platforms to realize the required $t_{2}^{AA}\neq t_{2}^{BB}$
asymmetry and thus the bowl-shaped $\mathbf{d}_\mathbf{k}$ geometry underlying
the QVH phase.

Beyond solid-state realizations, the QVH effect has also attracted extensive interest in metamaterial platforms, where QVH edge states have been observed and manipulated under highly controllable conditions~\cite{lu_observation_2017,dong_valley_2017,PhysRevLett.120.116802,barczyk_observation_2024}. Quantum-simulation platforms, in particular cold-atom systems, offer highly tunable Hamiltonians that are well suited to engineering such valley physics \cite{zhang2023tuning}. In cold-atom honeycomb optical lattices, state- and site-dependent laser-assisted tunneling can implement sublattice-asymmetric next-nearest-neighbor hoppings~\cite{jotzu2014experimental}, and BIS can be directly mapped by Bloch-state or quench tomography,
where the BIS is identified as the locus on which the time-averaged pseudospin component $d_z$ vanishes~\cite{yi2019observing,flaschner2016experimental,wu2016realization}. In optical-lattice systems, Bloch-state tomography~\cite{PhysRevResearch.5.L032016,doi:10.1126/science.aad4568} and time-of-flight imaging with Stern–Gerlach separation~\cite{valdes-curiel2021topological} allow direct measurement of the momentum distribution of each spin component, providing experimental access to the vector field $\mathbf{d}(\mathbf{k})/|\mathbf{d}(\mathbf{k})|$. In topo-electric circuits, the full $\mathbf{d}(\mathbf{k})$ is accessible
via the circuit admittance matrix \cite{PhysRevResearch.3.023056,lee2018topolectrical}. These platforms thus provide a natural
arena to visualize and test the complete sequence linking the bowl-shaped
$\mathbf{d}(\mathbf{k})$ geometry, the BIS formation, and
the emergence of valley edge modes predicted.

\textit{Conclusion}.—In conclusion, we have established a paradigm for realizing the intrinsic QVH effect within a monolayer system, eliminating the conventional requirement for domain walls. Our analysis reveals that the topological origin of this QVH phase corresponds to a specific geometric structure in parameter space: a bowl-shaped $\mathbf{d}_\mathbf{k}$ surface enclosing the origin. This unique topology guarantees the existence of valley-polarized edge states intrinsicly, providing a robust and compact mechanism for manipulating valley degrees of freedom in single-phase materials. 
Our findings fundamentally extend the conventional understanding 
of $\mathbf{d}(\mathbf{k})$-surface topology and
establishes a generic design principle that is readily applicable to diverse wave systems, from solid-state materials to photonic and acoustic metamaterials.

$\textit{Acknowledgment}s$.\textemdash
This work was financially supported
by the National Natural Science Foundation of China (Grants No. 124B2069, No. 12447146, No. 12447147, and No. 12374034),
the National Key R and D Program of China (Grant No. 2024YFA1409002),
the Quantum Science and Technology-National Science and Technology Major Project (Grant No. 2021ZD0302403), the China Postdoctoral Science Foundation (Grants No. 2025T180938 and No. 2024M760070) and the Postdoctoral Fellowship Program of
CPSF under Grant No. GZB20240031. We acknowledge the High-performance Computing Platform of Peking University
for providing computational resources.

\bibliography{ref}

\section*{End Matter}
\textit{Lattice-agnostic realization: square-lattice model}.—

\begin{figure}
\begin{centering}
\includegraphics[scale=0.35]{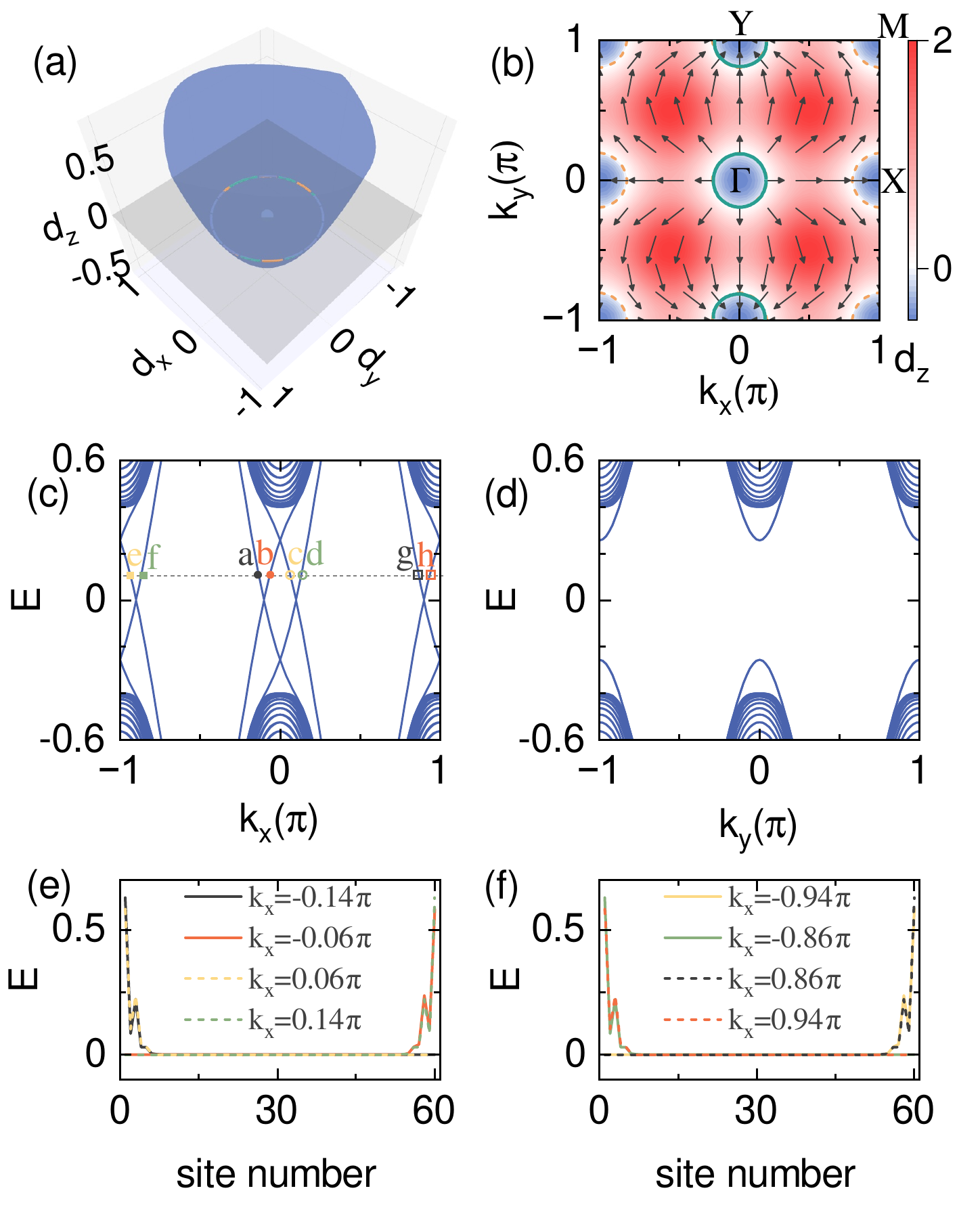}
\par\end{centering}
\caption{\label{fig:Square} (a) Bowl-shaped $\mathbf{d}_\mathbf{k}$ surface of the square-lattice model.
(b) Distribution of $\mathbf{d}(\mathbf{k})$ showing four valleys at $\Gamma$, $X$, $Y$, and $M$.
Four BIS loops overlap on the $\mathbf{d}_\mathbf{k}$ surface in (a) but separately wind around the four valleys in momentum space in (b).
(c) Ribbon spectrum along $x$: two pairs of co-propagating edge states appear near $k_x=0$ and $k_x=\pi$, corresponding to $C=+2$ and $C=-2$.
(d) Ribbon spectrum along $y$: no edge states appear due to opposite valley Chern numbers projecting onto the same momentum.
(e,f) Spatial distribution of the edge states for the points marked by colors in (c). We choose the model parameters as $A=1$, $m_1=0.6$, and $m=-0.4$.}
\end{figure}

To demonstrate that the mechanism based on a bowl-shaped $\mathbf{d}(\mathbf{k})$ geometry is not restricted to the honeycomb lattice, we construct an analogous realization on a square lattice described by a two-band Bloch Hamiltonian
\[
H_{\mathrm{sq}}(\mathbf{k}) = d_x(\mathbf{k})\sigma_x + d_y(\mathbf{k})\sigma_y + d_z(\mathbf{k})\sigma_z ,
\]
with
\[
\begin{aligned}
d_x &= A \sin k_x \cos k_y, \\
d_y &= A \sin k_y, \\
d_z &= 2 m_1 (\sin^2 k_x + \sin^2 k_y) + m .
\end{aligned}
\]
For representative parameters, the resulting $\mathbf{d}(\mathbf{k})$ surface exhibits a bowl-shaped curvature closely analogous to that of the honeycomb model, as shown in Fig.\ref{fig:Square}(a).

The corresponding $\mathbf{d}(\mathbf{k})$ texture over the Brillouin zone [Fig.\ref{fig:Square}(b)] reveals four inequivalent valleys located near $\Gamma$, $X$, $Y$, and $M$, each characterized by $(d_x,d_y)=(0,0)$ and acting as a source or sink of the in-plane field $\mathbf{d}_\perp=(d_x,d_y)$. The condition $d_z(\mathbf{k})=0$ defines four BIS loops, which overlap geometrically on the $\mathbf{d}_{\mathbf{k}}$ surface [see Fig.\ref{fig:Square}(a)] but remain spatially separated in momentum space [see Fig.\ref{fig:Square}(b)], each enclosing a single valley with alternating winding chirality.

This momentum-space separation has direct consequences for boundary spectra. For a ribbon finite along the $x$ direction, valleys with identical winding project onto the same one-dimensional momentum, giving rise to pairs of co-propagating edge modes near $k_x=0$ and $k_x=\pi$ [Fig.\ref{fig:Square}(c)]. In contrast, for a ribbon finite along $y$, valleys with opposite windings are projected onto the same momentum, leading to strong intervalley mixing and a fully gapped spectrum [Fig.\ref{fig:Square}(d)]. The spatial localization of the edge states [Fig.\ref{fig:Square}(e,f)] confirms their valley-polarized and topological origin.

Overall, this square-lattice realization demonstrates that the emergence of quantized valley edge transport is dictated by the global bowl-shaped geometry of the $\mathbf{d}(\mathbf{k})$ surface and the associated BIS winding. It therefore establishes this geometric mechanism as a lattice-agnostic route to intrinsic QVH physics.

\end{document}